\documentstyle[12pt,eepic,epic]{article}
\def\beq{\begin{equation}}
\def\eeq{\end{equation}}
\def\bea{\begin{eqnarray}}
\def\eea{\end{eqnarray}}
\def\nn{\nonumber}

\title{Test of the chiral structure and FCNC 
	in the quark sector by radiative $B$ meson decays}
\author{\vspace{1cm}\\
	{\bf L. T. Handoko} \thanks{On leave from P3FT-LIPI, Indonesia. 
	E-mail address : handoko@theo.phys.sci.hiroshima-u.ac.jp} \\
        Department of Physics, Hiroshima University \\
        1-3-1 Kagamiyama, Higashi Hiroshima  -  739, Japan\\
        \vspace{5mm}\\
	{\bf T. Yoshikawa} \thanks{
        E-mail address : yosikawa@theory.kek.jp}\\
	National Laboratory for High Energy Physics (KEK) \\
	Tsukuba - 305, Japan \\
	\vspace{3mm}\\}
\date{}

\begin{document}
\setlength{\baselineskip}{24pt}

\maketitle
\begin{picture}(0,0)
       \put(325,375){HUPD-9613}
       \put(325,360){KEK-TH-489}
       \put(325,345){August 1996}
       \put(250,325){revised version February 1997}
\end{picture}
\vspace{-24pt}

\setlength{\baselineskip}{8mm}
\thispagestyle{empty}
\renewcommand{\thesubsection}{\Roman{subsection}}

\begin{abstract}
        We study the effects of a vector-like SU(2) quark doublet   
        as a fourth generation. In this model 
	we examine the chiral structure and the FCNC in the 
	quark sector by using radiative $B$ meson decays in the allowed
        region for parameters from 
	$R_b = {\Gamma_{qq}}/{\Gamma_{\rm had.}}$. 
	We compute the ratio $R = {Br (b \rightarrow d \, \gamma)}/{
	Br (b \rightarrow s \, \gamma)}$ in the model 
	which realizes a different chiral 
	structure as well as FCNC. The constraints has been extracted 
	from the experimental results of $B$ meson decays, the $T_{\rm new}$ 
	parameter of oblique corrections and 
	$R_b$. Under the natural 
	assumption that the violation of the $V-A$ structure in the 
	light-quark sector is small, we can determine the allowed region 
	for most of the mixings parameters and the vector-like quark 
	masses.	We show that there will be significant deviations in $R$ 
	from the SM prediction due to the FCNC's and the violation of 
        the $V-A$ structure. 
\end{abstract}

\clearpage
\subsection{\bf INTRODUCTION}
\label{sec:intro}

The unitarity of the Cabibbo-Kobayashi-Maskawa (CKM) matrix 
and the universal $V-A$ structure of the interactions between the 
$SU(2)_L$ gauge bosons, quarks and leptons are two of the remarkable 
predictions of the $SU(2)_L \times U(1)_Y$ electroweak interactions. 
Deviations from these predictions would indicate new physics 
which should be directly detected in future high-energy experiments. 
At present, radiative $B$ decays provide an important test of the 
Standard Model (SM) and provide a sensitive probe of physics 
beyond the SM. The current result of the CLEO collaboration \cite{cleo95} 
gives a measurement of $Br(b \rightarrow s \, \gamma)$, 
and in the future it is 
expected that the $Br(b \rightarrow d \, \gamma)$ will also be 
determined. There is also an effort to search for flavor changing neutral 
currents (FCNC) in $B$ meson decays \cite{cleo96}. On the other hand, 
there were also discrepancies between the SM prediction for 
and experimental 
measurement of $R_b$, which is also a probe 
of new physics \cite{cornet} \footnote{The latest report 
\cite{aleph} point that the experimental result is consistent 
with the SM prediction. However the 1996 world average value of $R_b$
\cite{1996} 
is not still consistent.  
}. 
These experiments provide further motivation of this study. To explain the 
discrepancy many extensions of the SM have been presented. As the simplest 
extension of the SM, models that 
explain the data by using a new flavor mixing from a 
vector-like quark as a fourth
generation have been proposed by Ref. \cite{grimus,yoshikawa}. 
Here, we introduce an $SU(2)$ doublet of vector-like quarks in addition to 
the ordinary quarks in the SM. 
%
%
In the previous work of Ref. \cite{yoshikawa} one of the authors 
showed that there is an
allowed region for new flavor mixing between right handed third generation
and the fourth generation which is consistent with satisfies 
the measured value of $R_b$. 
In that case the mixing parameters must be large. 
If $R_b$ is consistent with the 
SM prediction, their region for large mixing remains.(See Fig. 1.) 
We interested in the new allowed 
region. This provides a real 
model for the deviation of $V-A$ chiral structure in the charged-current(CC) 
sector as pointed out in Ref. \cite{fujikawa}. At the same time, this 
also provides FCNC's in up and down quark sectors at once 
\cite{branco, lavoura}. In this work, we examine whether the region remains
after including  the constraints from radiative $B$ decays and 
oblique corrections.

Introducing a doublet vector-like quarks will violate the 
$V-A$ structure of the SM. This possible violation could provide useful 
information on new properties of the quarks at higher energy
scales as well as the origin of parity violation \cite{fujikawa}. 
Anyway, the induced FCNC due to the addition of the singlet or doublet 
vector-like quark has been studied by many authors, for example 
in \cite{branco, handoko}. In this paper, we study these FCNC's
 by using the ratio 
$R = {Br (b \rightarrow d \, \gamma)}/{Br (b \rightarrow s \, \gamma)}$ 
which is a sensitive probe to study the violation of CKM matrix as pointed 
out in \cite{gautam}. The constraints for the model are extracted from 
the experimental measurements of $b \rightarrow s \, \gamma$ process, 
the $T_{\rm new}$ parameter of oblique corrections and also 
$R_q = {\Gamma_{qq}}/{\Gamma_{\rm had.}}$. Further constraints will be 
obtained under the natural assumption that the violation of $V-A$ 
structure in the light-quarks sector is small.

This paper is organized as follows. In Sec.\ref{sec:model}
we briefly describe 
the model and show how the violation of the $V-A$ structure and 
FCNC's appear. In Sec.\ref{Rb}, 
we review the constraints from $R_b$ on the mixing parameters. 
In Sec.\ref{sec:radiative},
we describe the radiative $B$ decays in the model. 
In Sec.\ref{sec:constraint}, some constraints 
from the experimental results for the vector-like quark masses, the size 
of FCNC and the CKM matrix for right-handed quarks 
are given. Before going to the 
conclusions, in Sec.\ref{sec:result} the numerical results which show 
the effects on $R$ are presented.

\subsection{\bf THE MODEL}
\label{sec:model}

We study the SM extended with an $SU(2)$ vector-like quark doublet
as a fourth generation. In this extension, the charge assignments 
of the quark sector under 
the electroweak symmetry $SU(2)_L \otimes U(1)_Y $ are, 
\begin{center}
\begin{tabular}{lclcl}
	${\tilde{Q}_L}^i$ 	& = & $\left(
			\begin{array}{c}
				\tilde{u}^i \\
				\tilde{d}^i \\
    			\end{array}
                	\right)_L$ 
		& $\rightarrow$ & $\left( 2, \frac{1}{3}\right)$ \, , \\
	${\tilde{u}_R}^i$ 	& & 
		& $\rightarrow$ & $\left( 1, \frac{4}{3}\right)$ \, , \\
	${\tilde{d}_R}^i$ 	& & 
		& $\rightarrow$ & $\left( 1, -\frac{2}{3}\right)$ \, , \\
	${\tilde{Q}_{L/R}}^4$ & = & $\left(
			\begin{array}{c}
				\tilde{u}^4 \\
				\tilde{d}^4 \\
    			\end{array}
                	\right)_{L/R}$ 
		& $\rightarrow$ & $\left( 2, \frac{1}{3}\right)$ \, .
\end{tabular}\\
\end{center}
Because the vector-like quarks are an $SU(2)$ doublet, the left-handed 
and right-handed chiralities have the same charge under an $SU(2)_L$ 
transformation. The tilde on the fields denotes weak-eigenstate 
quark fields, $i$ denotes the flavor, and, as usual the chiralities are 
$Q_{L/R} \equiv {(1 \mp \gamma_5)}Q/2$.

There is an ambiguity in reproducing the masses of the vector-like 
quarks, but the study should not be altered with the choice. 
In this paper we adopt the same procedure and notations 
as Ref.\cite{handoko}. 
We consider one Higgs doublet which is the same as in the SM 
and introduce 
explicit bare-mass terms for the vector-like quarks. 
Then, the Lagrangian for the 
Yukawa sector becomes,
\begin{equation}
	{\cal L}_Y = 
		- {f_d}^{\alpha i} \, \bar{\tilde{Q}_L}^\alpha \, 
			\phi \, {\tilde{d}_R}^i 
		- {f_u}^{\alpha i} \, \bar{\tilde{Q}_L}^\alpha \, 
			\tilde{\phi} \, {\tilde{u}_R}^i 
		- f^{\alpha 4} \, v^\prime \, \bar{\tilde{Q}_L}^\alpha \, 
			{\tilde{Q}_R}^4 
		+ {\rm h.c.} \; ,
	\label{eqn:lagyukawa}
\end{equation}
with $i = 1, 2, 3$ and $\alpha = 1, 2, 3, 4$. In order to obtain  
the masses and mass eigenstates, we must diagonalize 
the Yukawa couplings above as 
follows: 
\bea
    D_L \pmatrix{ f_d^{\alpha i}\frac{v}{\sqrt{2}}\, ,
                              & f^{\alpha 4}v^\prime} D_R^\dagger
        &=& m_d, \\
    U_L \pmatrix{ f_u^{\alpha i}\frac{v}{\sqrt{2}}\, ,
                              & f^{\alpha 4}v^\prime} U_R^\dagger
        &=& m_u,
\eea
%
where $m_d$ and $m_u$ are the diagonalized mass matrices and 
$D_{L/R}$ and $U_{L/R}$ are $4 \times 4$ unitary matrices 
which relate the weak-eigenstates to the mass-eigenstates as, 
\begin{eqnarray}
	{\tilde{d}_{L/R}}^\alpha & \equiv & 
		{D_{L/R}}^{\alpha \beta} \, {d_{L/R}}^\beta \; , 
		\label{eqn:tilded} \\ 
	{\tilde{u}_{L/R}}^\alpha & \equiv & 
		{U_{L/R}}^{\alpha \beta} \, {u_{L/R}}^\beta \; ,
		\label{eqn:tildeu}
\end{eqnarray}
where $\alpha$ and $\beta$ denote the flavor for four generations. 
Note that $f^{\alpha 4}$ in both the up and the down sectors 
must have same value
in the third term of Eq. (1), because the term show 
the $SU(2)$ invariant mass 
of vector-like doublet quark. 
Hence there is a relation between the up and the down sector
as follows:
\bea
         f^{\alpha 4} \, v^\prime & \equiv & 
		({D_L}^\dagger)^{\alpha \beta} \, 
		{m_d}^\beta \, {D_R}^{\beta 4} 
	      = ({U_L}^\dagger)^{\alpha \beta} \, 
		{m_u}^\beta \, {U_R}^{\beta 4} \; .
			\label{eqn:f} 
\eea  

Here, we give a brief description of how 
the violation of the $V-A$ structure is realized. As an example, we show  
only the $W^\pm$ and $Z$ sectors. 
The original interactions in the weak-eigenstates 
are, 
\bea
        {\cal L}_{Z} &=& \frac{g}{2 \cos \theta_W }
               \left[
                  \bar{\tilde{u}}_L^\alpha \gamma^\mu
                    \left(
                       1 - \frac{4}{3}\sin^2 \theta_W 
                                                   \right)
                                       \tilde{u}_L^\alpha  
           +      \bar{\tilde{u}}_R^i \gamma^\mu
                    \left(
                       - \frac{4}{3} \sin^2 \theta_W 
                                                   \right) 
                                       \tilde{u}_R^i \right.\nn \\
       &\, &  \hspace{6.2cm} +       \bar{\tilde{u}}_R^4 \gamma^\mu
                    \left(
                       1 - \frac{4}{3}\sin^2 \theta_W 
                                                   \right) 
                                       \tilde{u}_R^4 \nn \\
       &\, &  \hspace{1.2cm}  + \,
                  \bar{\tilde{d}}_L^\alpha \gamma^\mu
                    \left(
                       - 1 + \frac{2}{3}\sin^2 \theta_W 
                                                   \right) 
                                       \tilde{d}_L^\alpha  
           +      \bar{\tilde{d}}_R^i \gamma^\mu
                    \left(
                        \frac{2}{3}\sin^2 \theta_W 
                                                   \right) 
                                       \tilde{d}_R^i  \\
       &\, & \left. \hspace{6.2cm}  +       \bar{\tilde{d}}_R^4 \gamma^\mu
                    \left(
                       - 1 + \frac{4}{3}\sin^2 \theta_W 
                                                   \right) 
                                       \tilde{d}_R^4 
                                                         \right] Z_\mu
                                                       ,\nn \\[5mm]
	{\cal L}_{W^\pm} &=& 
		\frac{g}{\sqrt{2}} \, \left( 
		\bar{\tilde{u}}_L^\alpha \, \gamma^\mu \, {\tilde{d}}_L^\alpha 
		+ \bar{\tilde{u}}_R^4 \, \gamma^\mu \, {\tilde{d}}_R^4 
		\right) \, W^\pm_\mu + {\rm h.c.} \: ,
		\label{eqn:lagwwe} 
\eea
because the left and right handed fourth generation quarks are in an 
$SU(2)$ doublet. The transformation into the mass-eigenstates 
gives, 
\bea
        {\cal L}_{Z} &=& \frac{g}{2 \cos \theta_W }
               \left[
                  \bar{{u}}_L^\alpha \gamma^\mu 
                    \left(
                       1 - \frac{4}{3}\sin^2 \theta_W 
                                                   \right)
                                       {u}_L^\alpha  
           +      \bar{{u}}_R^\alpha 
                    \left(\gamma^\mu 
                       - \frac{4}{3} \sin^2 \theta_W 
                                                   \right) 
                                       {u}_R^\alpha \right.\nn \\
       &\, &  \hspace{6.2cm} +       \bar{{u}}_R^\alpha \gamma^\mu 
                    \left( U_R^{\alpha 4} {U_R^\dagger}^{4 \beta}
                                                           \right) 
                                       {u}_R^\beta \nn \\
       &\, &  \hspace{1.2cm}  + \,
                  \bar{{d}}_L^\alpha \gamma^\mu 
                    \left(
                       - 1 + \frac{2}{3}\sin^2 \theta_W 
                                                   \right) 
                                       {d}_L^\alpha  
           +      \bar{{d}}_R^\alpha \gamma^\mu 
                    \left(
                        \frac{2}{3}\sin^2 \theta_W 
                                                   \right) 
                                       {d}_R^\alpha \nn \\
       &\, & \left. \hspace{6.2cm}  +  \bar{{d}}_R^\alpha \gamma^\mu 
                    \left(
                       - D_R^{\alpha 4} {D_R^\dagger}^{4 \beta} 
                                                   \right) 
                                       {d}_R^\beta 
                                                         \right] Z_\mu,\\[5mm]
	{\cal L}_{W^\pm} &=&  
		\frac{g}{\sqrt{2}} \, \left[ 
		\bar{u_L}^\alpha \, \gamma^\mu \, \left( 
			{U_L}^{\alpha \delta} \, {D^\dagger_L}^{\delta \beta} 
			\right) \, {d_L}^\beta 
		+ \bar{u_R}^\alpha \, \gamma^\mu \, \left( 
			{U_R}^{\alpha 4} \, {D^\dagger_R}^{4 \beta} 
			\right) \, {d_R}^\beta 
		\right] \, W^+_\mu + {\rm h.c.} \: ,\nn \\
		\label{eqn:lagwme} 
\eea
where we have used Eqs. (\ref{eqn:tilded}) and (\ref{eqn:tildeu}).  
The vector-like doublet quark induces 
FCNC in the up and down right-handed quark sectors, 
which for convenience we describe in the following way, 
\begin{eqnarray}
	{z_d}^{\alpha \beta} & \equiv & 
		\sum_{i = 1}^3 \, {D_R}^{\alpha i} \, {D_R^\dagger}^{i \beta}
		= \delta^{\alpha \beta} - {D_R}^{\alpha 4} \, 
 		{D_R^\dagger}^{4 \beta} \; , 
		\label{eqn:zd} \\
	{z_u}^{\alpha \beta} & \equiv & 
		\sum_{i = 1}^3 \, {U_R}^{\alpha i} \, {U_R^\dagger}^{i \beta}
		= \delta^{\alpha \beta} - {U_R}^{\alpha 4} \, 
			{U_R^\dagger}^{4 \beta} \; , 
		\label{eqn:zu} 
\end{eqnarray}
by using unitarity of $U_{L/R}$ and $D_{L/R}$ matrices. 
$z_q$ ($q = u, d$) indicates the size of the induced FCNC's \cite{branco}. 
In the case where $\alpha = \beta$, one finds 
${z_d}^{\alpha \alpha} = 1 - \left| {D_R}^{\alpha 4} \right|^2 $ and 
${z_u}^{\alpha \alpha} = 1 - \left| {U_R}^{\alpha 4} \right|^2 $. 
On the other hand, in the CC sector the right CKM matrix includes 
new effects.
${U_L}^{\alpha \delta} \, {V^\dagger_L}^{\delta \beta}$ and 
${U_R}^{\alpha 4} \, {V^\dagger_R}^{4 \beta}$ show
the left and right CKM matrices in Eq. (\ref{eqn:lagwme}). 
\begin{eqnarray}
	{V_{LCKM}}^{\alpha \beta} & \equiv & 
		\sum_{\rho = 1}^4 \, {U_L}^{\alpha \rho} \, 
			{D_L^\dagger}^{\rho \beta} \; , 
		\label{eqn:vlckm} \\
	{V_{RCKM}}^{\alpha \beta} & \equiv & 
		{U_R}^{\alpha 4} \, 
			{D_R^\dagger}^{4 \beta} \; . 
		\label{eqn:vrckm} 
\end{eqnarray}

We use the ratio of the 
elements of left and right CKM matrix
for convenience to make comparison with the SM's one and give 
the bounds from the experimental results. The ratio must be defined as
\begin{eqnarray}
	V^{\alpha \beta} & \equiv &
		\frac{{V_{RCKM}}^{\alpha \beta}}{
			{V_{LCKM}}^{\alpha \beta}} 
		\nonumber \\
	& = & \left| \frac{{V_{RCKM}}^{\alpha \beta}}{
		{V_{LCKM}}^{\alpha \beta}} \right| 
		\, e^{i \theta_{\alpha \beta}}\; , 
	\label{eqn:phasevrvl}
\end{eqnarray}
where 
\begin{equation}
	\theta_{\alpha \beta} \equiv 
		{\rm arg} \left( \frac{{V_{RCKM}}^{\alpha \beta}}{
		{V_{LCKM}}^{\alpha \beta}} \right) \; .
	\label{eqn:thetaalbe}
\end{equation}
This phase factors should induce new $CP$ violation sources 
in the model as generally a $4 \times 4$ matrix has three phases. 
$\left| V^{\alpha \beta} \right|$ indicates the size of 
the violation of the $V-A$ structure in the theory. Note that in general 
$V_{RCKM}$ is not a unitary matrix, while $V_{LCKM}$ retains its unitarity.
This point is differs from the singlet vector-like quark model 
\cite{branco, handoko} which violates the unitarity of $V_{LCKM}$.

Substituting the above relations into Eq. (\ref{eqn:lagyukawa}), 
the Lagrangian for the neutral and charged Higgs sector becomes, 
\begin{eqnarray}
	{\cal L}_{H,\chi^0} & = & 
	      	-\frac{g}{2 \, M_W} \, \bar{q}^\alpha \, 
		{z_q}^{\alpha \beta} \, m_{q}^\beta \, R \, q^\beta 
		\left( H \mp i \, \chi^0 \right) 
		+ {\rm h.c.} \; , 
		\label{eqn:lagphi} \\
	{\cal L}_{\chi^\pm} & = & 
		\frac{g}{\sqrt{2} \, M_W} \, 
		\bar{u}^\alpha \, {V_{LCKM}}^{\alpha \beta} 
			\left[ 
			\left( 1 - V^{\alpha \beta} 
			\frac{m_{d}^\beta}{m_{u}^\alpha} \right) \, 
			m_{u}^\alpha \, L 
		\right. \nonumber \\
	& & \left. \; \; \; \; \; \; \;  \; \; \; \; \; \; \; \; \; \; \;  
			\; \; \; \; \; \; \;  \; \; \; \; \; \; \; \; \; 
		      - \left( 1 - V^{\alpha \beta} 
			\frac{m_{u}^\alpha}{m_{d}^\beta} \right) \, 
			m_{d}^\beta \, R 
			\right] 
		d^\beta \, \chi^+ + {\rm h.c.} \; , 
		\label{eqn:lagchi} 
\end{eqnarray}
where the upper sign is for $q = u$ and the lower one is for $q = d$. 
The same procedure gives the Lagrangian for the gauge boson quark sector, 
\begin{eqnarray}
	{\cal L}_A & = & g \, \sin \theta_W \, Q_q \, 
		\bar{q}^\alpha \, \gamma_\mu \, q^\alpha \, A^\mu \; , 
		\label{eqn:lagphoton} \\
	{\cal L}_Z & = & \pm \frac{g}{2 \, \cos \theta_W} \, 
		\bar{q}^\alpha \, \gamma_\mu \, \left[ 
			g_q \, \delta^{\alpha \beta} \, L 
			\pm \left( g_q \, \delta^{\alpha \beta} 
			- {z_q}^{\alpha \beta} \right) \, R \right] \, 
			q^\beta \, Z^\mu \; , 
		\label{eqn:lagz} \\
	{\cal L}_{W^\pm} & = & 
		\frac{g}{\sqrt{2}} \, \bar{u}^\alpha \, 
		{V_{LCKM}}^{\alpha \beta} \, \gamma_\mu \, \left( 
			L + V^{\alpha \beta} \, R \right) \, 
		d^\beta \, {W^\mu}^+ + {\rm h.c.} \; . 
		\label{eqn:lagw} 
\end{eqnarray}
Here $\theta_W$ is the Weinberg angle, 
$g_q \equiv \left( 1 \mp 2 \, Q_q \, \sin^2 \theta_W \right)$, 
and $Q_q$ is the electric charge ($Q_u = 2/3$ and $Q_d = -1/3$). 
Therefore, in the present model, the violation of the $V-A$ structure appears
in the CC sector (Eqs. (\ref{eqn:lagchi}) 
and (\ref{eqn:lagw})), while the FCNC's appear in the neutral-current(NC) 
sector (Eqs. (\ref{eqn:lagphi}) and (\ref{eqn:lagz})). It is clear that 
when one puts $V^{\alpha \beta} = 0$ and 
${z_q}^{\alpha \beta} = \delta^{\alpha \beta}$, the SM Lagrangian 
is restored. 

In this model there are many newly introduced parameters. Hence, 
for simplicity, we make the following assumptions: 
(A) the SM-like $3 \times 3$ matrix is retained for 
the left-handed CKM matrix, 
{\it i.e.}, 
\begin{equation}
	V_{LCKM} \sim \left(
	\begin{array}{cc}
		V_{CKM}	& -\times \\
		\times	& 1
	\end{array}
	\right) \; , 
	\label{eqn:vlckma}
\end{equation}
and (B) there are right-handed mixings
among the heavy quarks but not among the light quarks. 
This can be realized by using a special form of the unitary matrices $U_R$ 
and $D_R$ as follows:
\begin{equation}
	\sim \left(
		\begin{array}{cccc}
		1      & \times & \times & \times \\
		\times & 1	& \times & \times \\
		\times & \times	& \cos \theta_{U/D} & -\sin \theta_{U/D} \\
		\times & \times	& \sin \theta_{U/D} & \cos \theta_{U/D} 
		\end{array}
	\right) \; , 
	\label{eqn:ass}
\end{equation}
which permits possible large mixings between the third and fourth 
generations of right-handed quarks. `$\times$' denotes a small ($\ll 1$) 
but non-zero element.

\subsection{CONSTRAINT FROM $R_b$}
\label{Rb}
Ref.\cite{yoshikawa} showed that there are 
regions allowed by the measurement of $R_b$ for nonzero values of 
the mixing parameters between the third and the fourth generations. 
The value of the parameter 
$\mid D_R^{34} \mid^2$ had to be larger than $4 \sin^2 \theta_W /3$.

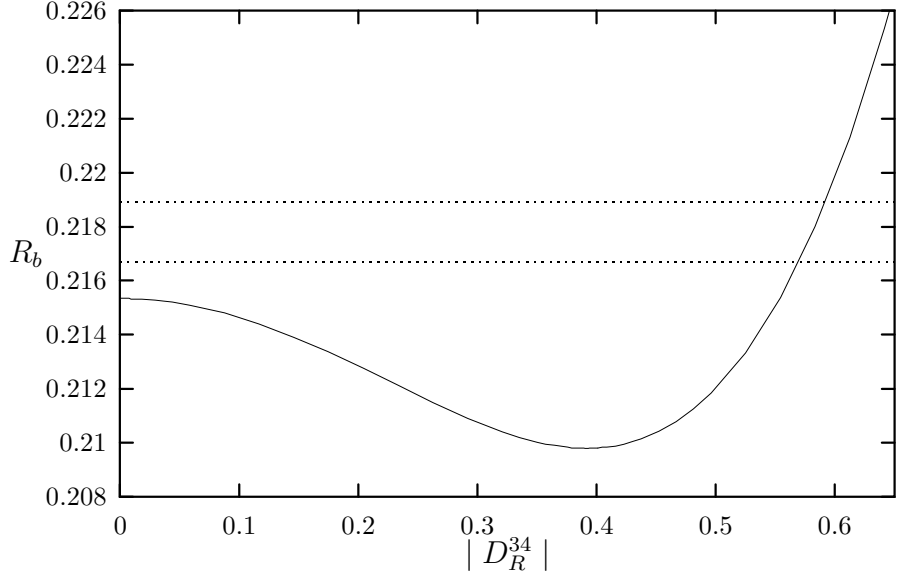
\begin{figure}[t]
        \begin{center}
\setlength{\unitlength}{0.240900pt}
\begin{picture}(1500,900)(0,0)
\tenrm
\thicklines \path(220,113)(240,113)
\thicklines \path(1436,113)(1416,113)
\put(198,113){\makebox(0,0)[r]{0.208}}
\thicklines \path(220,198)(240,198)
\thicklines \path(1436,198)(1416,198)
\put(198,198){\makebox(0,0)[r]{0.21}}
\thicklines \path(220,283)(240,283)
\thicklines \path(1436,283)(1416,283)
\put(198,283){\makebox(0,0)[r]{0.212}}
\thicklines \path(220,368)(240,368)
\thicklines \path(1436,368)(1416,368)
\put(198,368){\makebox(0,0)[r]{0.214}}
\thicklines \path(220,453)(240,453)
\thicklines \path(1436,453)(1416,453)
\put(198,453){\makebox(0,0)[r]{0.216}}
\thicklines \path(220,537)(240,537)
\thicklines \path(1436,537)(1416,537)
\put(198,537){\makebox(0,0)[r]{0.218}}
\thicklines \path(220,622)(240,622)
\thicklines \path(1436,622)(1416,622)
\put(198,622){\makebox(0,0)[r]{0.22}}
\thicklines \path(220,707)(240,707)
\thicklines \path(1436,707)(1416,707)
\put(198,707){\makebox(0,0)[r]{0.222}}
\thicklines \path(220,792)(240,792)
\thicklines \path(1436,792)(1416,792)
\put(198,792){\makebox(0,0)[r]{0.224}}
\thicklines \path(220,877)(240,877)
\thicklines \path(1436,877)(1416,877)
\put(198,877){\makebox(0,0)[r]{0.226}}
\thicklines \path(220,113)(220,133)
\thicklines \path(220,877)(220,857)
\put(220,68){\makebox(0,0){0}}
\thicklines \path(407,113)(407,133)
\thicklines \path(407,877)(407,857)
\put(407,68){\makebox(0,0){0.1}}
\thicklines \path(594,113)(594,133)
\thicklines \path(594,877)(594,857)
\put(594,68){\makebox(0,0){0.2}}
\thicklines \path(781,113)(781,133)
\thicklines \path(781,877)(781,857)
\put(781,68){\makebox(0,0){0.3}}
\thicklines \path(968,113)(968,133)
\thicklines \path(968,877)(968,857)
\put(968,68){\makebox(0,0){0.4}}
\thicklines \path(1155,113)(1155,133)
\thicklines \path(1155,877)(1155,857)
\put(1155,68){\makebox(0,0){0.5}}
\thicklines \path(1342,113)(1342,133)
\thicklines \path(1342,877)(1342,857)
\put(1342,68){\makebox(0,0){0.6}}
\thicklines \path(220,113)(1436,113)(1436,877)(220,877)(220,113)
\put(45,495){\makebox(0,0)[l]{\shortstack{$ R_b $}}}
\put(828,23){\makebox(0,0){$ \mid D_{R}^{34}\mid $}}
\thinlines \path(220,425)(220,425)(222,425)(223,425)(225,425)
(227,425)(229,425)(230,425)(234,425)(237,424)(240,424)(247,424)
(254,424)(261,423)(275,422)(288,421)(302,419)(329,414)(384,402)
(438,385)(493,364)(547,341)(602,315)(657,288)(711,261)(766,236)
(820,215)(847,206)(875,199)(888,196)(902,194)(916,192)(923,191)
(929,190)(936,190)(940,190)(941,190)(943,190)(945,190)(946,190)
(948,190)(950,189)(952,189)(953,189)(955,189)(957,190)(958,190)
(960,190)(963,190)(967,190)(970,190)
\thinlines \path(970,190)(977,191)(984,191)(998,193)(1011,196)
(1025,200)(1038,204)(1066,216)(1093,231)(1120,251)(1148,276)
(1202,339)(1257,426)(1311,538)(1366,678)(1420,849)(1428,877)
\thinlines \dashline[-10]{3}(220,576)(220,576)(232,576)(245,576)
(257,576)(269,576)(281,576)(294,576)(306,576)(318,576)(331,576)
(343,576)(355,576)(367,576)(380,576)(392,576)(404,576)(417,576)
(429,576)(441,576)(453,576)(466,576)(478,576)(490,576)(503,576)
(515,576)(527,576)(539,576)(552,576)(564,576)(576,576)(588,576)
(601,576)(613,576)(625,576)(638,576)(650,576)(662,576)(674,576)
(687,576)(699,576)(711,576)(724,576)(736,576)(748,576)(760,576)
(773,576)(785,576)(797,576)(810,576)(822,576)
\thinlines \dashline[-10]{3}(822,576)(834,576)(846,576)(859,576)
(871,576)(883,576)(896,576)(908,576)(920,576)(932,576)(945,576)
(957,576)(969,576)(982,576)(994,576)(1006,576)(1018,576)(1031,576)
(1043,576)(1055,576)(1068,576)(1080,576)(1092,576)(1104,576)
(1117,576)(1129,576)(1141,576)(1153,576)(1166,576)(1178,576)
(1190,576)(1203,576)(1215,576)(1227,576)(1239,576)(1252,576)
(1264,576)(1276,576)(1289,576)(1301,576)(1313,576)(1325,576)
(1338,576)(1350,576)(1362,576)(1375,576)(1387,576)(1399,576)
(1411,576)(1424,576)(1436,576)
\thinlines \dashline[-10]{3}(220,482)(220,482)(232,482)(245,482)
(257,482)(269,482)(281,482)(294,482)(306,482)(318,482)(331,482)
(343,482)(355,482)(367,482)(380,482)(392,482)(404,482)(417,482)
(429,482)(441,482)(453,482)(466,482)(478,482)(490,482)(503,482)
(515,482)(527,482)(539,482)(552,482)(564,482)(576,482)(588,482)
(601,482)(613,482)(625,482)(638,482)(650,482)(662,482)(674,482)
(687,482)(699,482)(711,482)(724,482)(736,482)(748,482)(760,482)
(773,482)(785,482)(797,482)(810,482)(822,482)
\thinlines \dashline[-10]{3}(822,482)(834,482)(846,482)(859,482)
(871,482)(883,482)(896,482)(908,482)(920,482)(932,482)(945,482)
(957,482)(969,482)(982,482)(994,482)(1006,482)(1018,482)(1031,482)
(1043,482)(1055,482)(1068,482)(1080,482)(1092,482)(1104,482)
(1117,482)(1129,482)(1141,482)(1153,482)(1166,482)(1178,482)
(1190,482)(1203,482)(1215,482)(1227,482)(1239,482)(1252,482)
(1264,482)(1276,482)(1289,482)(1301,482)(1313,482)(1325,482)
(1338,482)(1350,482)(1362,482)(1375,482)(1387,482)(1399,482)
(1411,482)(1424,482)(1436,482)
\end{picture}
        \end{center}
	\caption{$R_b$ as a function of $\mid D_R^{34}\mid$ }
	\label{fig:rb}
\end{figure}

$R_b$ is defined as 
$R_b \equiv \Gamma (Z \rightarrow b \, \bar{b})/ 
\Gamma (Z \rightarrow {\rm hadrons})$. An estimate for the 
upper-bound of $\left| {D_R}^{34} \right|$ can be extracted
\cite{grimus, yoshikawa}, that is
\begin{equation}
	R_b = \frac{\Gamma_{bb}}{\Gamma_{uu} + \Gamma_{cc} + 
		\Gamma_{dd} + \Gamma_{ss} + \Gamma_{bb}} \; ,
		\label{eqn:rb}
\end{equation}
where 
\bea
	\Gamma_{q^iq^i} &\equiv& 
	\left( {g_q^i} + {\delta_{SM}}^{q^i q^i} \right)^2 + 
	\left( g_q^i - {z_q}^{q^i q^i} \right)^2,   
	\label{eqn:gammaqq}
\eea
with ${\delta_{SM}}^{q^i q^i}$ is the one-loop corrections of 
$Z \rightarrow q^i \, \bar{q}^i$ within the SM, and all of the fermion 
masses are neglected. $\left| {D_R}^{34} \right|$ 
appears in eqs.(25) through ${z_d}^{bb}$ 
according to Eqs. (\ref{eqn:zd}). Since 
${\delta_{SM}}^{q^i q^i}$'s are insignificant except for $q^i = b$, 
in the numerical calculations we will keep the one-loop correction only 
in $\Gamma_{bb}$ \cite{cornet}. From Eq. (\ref{eqn:gammaqq}),
\bea
  \Gamma_{bb} = \left( 1 - \frac{2}{3} \sin^2 \theta_W 
                                + {\delta_{SM}}^{bb}\right)^2 
              + \left( - \frac{2}{3} \sin^2 \theta_W +  
                       \left| {D_R}^{34} \right|^2 \right)^2 . 
\eea
In Fig. \ref{fig:rb} $R_b$ is shown  as a function of 
$\left| {D_R}^{34} \right|$ for $R_c = 0.172$ 
and ${\delta_{SM}}^{bb} = - 0.011$.
From the figure, we find the parameter $\left| {D_R}^{34} \right|$  
must be larger than 0.55 
when the measurement values of $R_b$ is larger than the SM prediction.  
For $R_b = 0.2178 \pm 0.0011$ and $R_c = 0.172$ \cite{1996} the bound is 
\beq
	0.57 < \left| {D_R}^{34} \right| < 0.59 \; . 
\label{Dbound}
\eeq
Even if the experiment of $R_b$ is consistent with the SM, the figure 
shows that the 
region for large value of $\left| {D_R}^{34} \right|$ may be survive.

However, if $\left| {D_R}^{34} \right|$ has a large value, 
from Eq. (\ref{eqn:vrckm}), $V_{RCKM}^{33}$ may also be large, or 
$U_R^{34}$ may be very small.   

\subsection{\bf RADIATIVE $B$ DECAYS}
\label{sec:radiative}

Next, We consider radiative $B$ decays in the 
model. 
After calculating the relevant Feynman diagrams including the neutral Higgs 
contributions, we find the amplitude for on-shell 
$B \rightarrow X_{d_l} \, \gamma$ up to second order in the external 
momenta as below, 
\bea
	T &=& \frac{G_F \, e}{8 \sqrt{2} \, \pi^2} \, 
		{V_{LCKM}^\ast}^{t d_l} \, {V_{LCKM}}^{tb} \, Q_u \, 
		\bar{d}_l (p^\prime) \nn \\
    \ &\ & \hspace{2cm} \times                
                \left[ \gamma_\mu, \gamma_\nu {q}^\nu \right] \, 
		\left( F_L(m_{d_l}) \, m_{d_l} \, L + F_R(m_{d_l}) \, m_b \, R 
		\right) \, b (p) \, \epsilon^\mu \; , 
		\label{eqn:t}
\eea
with  
\begin{eqnarray}
	F_L(m_{d_l}) & = & 
		\sum_{i = t, u^4} \, 
			\frac{{V_{LCKM}^\ast}^{i d_l} \, {V_{LCKM}}^{ib}}{
			{V_{LCKM}^\ast}^{t d_l} \, {V_{LCKM}}^{tb}} \, \left[ 
		\left( 1 + {V^\ast}^{i d_l} \, V^{ib} 
			\frac{m_b}{m_{d_l}} \right) F_1 (x_i) 
		\right. \nonumber \\
	& & \left. \; \; \; \; \; \; \;  \; \; \; \; \; \; \; \; \; \; \;  
			\; \; \; \; \; \; \;  \; \; \; \; \; \; \; \; \; 
			+ {V^\ast}^{i d_l} \, F_2 (x_i, m_{d_l}) 
			+ \left( 1 - {V^\ast}^{i d_l} \frac{m_i}{m_{d_l}} 
			\right) F_3 (x_i) \right] 
		\nonumber \\ 
	& & 	- \frac{Q_d}{Q_u} \, \frac{{z_d}^{d_l b}}{
			{V_{LCKM}^\ast}^{t d_l} \, {V_{LCKM}}^{tb}}
			\, \left[ g_d \, F_4 (r_{d_l}) + 
                        \left| {D_R}^{4 4} \right|^2 \, \left(
			F_5 (r_{d^4}, w_{d^4}) + F_6 (r_{d^4})
			\right) \right. \nonumber \\
	& & \left. \; \; \; \; \; \; \;  \; \; \; \; \; \; \; \; \; \; \;  
			\; \; \; \; \; \; \;  \; \; \; \; \; \; \; \; \; 
		+ \frac{1}{3} \, \left( \left| {D_R}^{14} \right|^2 
		+ \left| {D_R}^{24} \right|^2 + \left| {D_R}^{34} \right|^2
		\right) \right] \; , 
		\label{eqn:fl} \\
	F_R(m_{d_l}) & = & 
		\sum_{i = t, u^4} \, 
			\frac{{V_{LCKM}^\ast}^{i d_l} \, {V_{LCKM}}^{ib}}{
			{V_{LCKM}^\ast}^{t d_l} \, {V_{LCKM}}^{tb}} \, \left[ 
		\left( 1 + {V^\ast}^{i d_l} \, V^{ib} 
			\frac{m_{d_l}}{m_b} \right) F_1 (x_i) 
		\right. \nonumber \\
	& & \left. \; \; \; \; \; \; \;  \; \; \; \; \; \; \; \; \; \; \;  
			\; \; \; \; \; \; \;  \; \; \; \; \; \; \; \; \; 
			+ V^{ib} \, F_2 (x_i, m_b) 
			+ \left( 1 - V^{ib} \frac{m_i}{m_b} 
			\right) F_3 (x_i) \right]
		\nonumber \\ 
	& & 	- \frac{Q_d}{Q_u} \, \frac{{z_d}^{d_l b}}{
			{V_{LCKM}^\ast}^{t d_l} \, {V_{LCKM}}^{tb}}
			\, \left[ g_d \, F_4 (r_b) + 
			\left| {D_R}^{44} \right|^2 \, \left(
			F_5 (r_{d^4}, w_{d^4}) + F_6 (r_{d^4}) 
			\right) \right. \nonumber \\
	& & \left. \; \; \; \; \; \; \;  \; \; \; \; \; \; \; \; \; \; \;  
			\; \; \; \; \; \; \;  \; \; \; \; \; \; \; \; \; 
		+ \frac{1}{3} \, \left( \left| {D_R}^{14} \right|^2 
		+ \left| {D_R}^{24} \right|^2 + \left| {D_R}^{34} \right|^2
		\right) \right] \; , 
		\label{eqn:fr}
\end{eqnarray}
where $x_i \equiv ({m_i}/{M_W})^2$, $r_\alpha \equiv ({m_\alpha}/{M_Z})^2$  
and $w_\alpha \equiv ({m_\alpha}/{M_H})^2$. For simplicity  
we have employed the relations 
\begin{equation}
	{z_d^\dagger}^{d_l \alpha} \, {z_d}^{\alpha b} \simeq \left\{
	\begin{array}{ll}
		0 	& {\rm for} \; \; \alpha = d_l, b \\
		-{z_d}^{d_l b} \, \left| {D_R}^{\alpha 4} \right|^2 & 
			{\rm for} \; \; \alpha \neq d_l, b 
	\end{array} \right.
	\label{eqn:zddlb}
\end{equation}
by using Eq. (\ref{eqn:zd}) and the approximation that 
$r_{d_i} \simeq w_{d_i} \simeq 0$. Then $F_5(r_{d^i}, w_{d^i}) \sim 0$ and 
$F_6(r_{d^i}) \sim 1/3$. On the other hand, $F_5(r_{d^4}, w_{d^4})$ and 
$F_6(r_{d^4})$ could be significant. One assumption here is that 
${V_{RCKM}}^{u^i d^j}$ for $u^i = u, c$ and $d^j = d_l, b$ are negligible 
compared with the other elements. This assumption and how it is realized
will 
be discussed in detail in the next section. On the other hand, since 
the unitarity of $V_{LCKM}$ is still retained, the relation 
\begin{equation}
	{V_{LCKM}^\ast}^{u d_l} {V_{LCKM}}^{ub} + 
	{V_{LCKM}^\ast}^{c d_l} {V_{LCKM}}^{cb}
	= -{V_{LCKM}^\ast}^{t d_l} {V_{LCKM}}^{tb} -
	{V_{LCKM}^\ast}^{u^4 d_l} {V_{LCKM}}^{u^4 b}
	\label{eqn:vlckmunit}
\end{equation}
is valid. Similar to Eq. (\ref{eqn:phasevrvl}), there are also 
phase factors which induce $CP$ violation from the mixings of FCNC, 
that is 
\begin{equation}
	\frac{{z_d}^{d_l b}}{{V_{LCKM}^\ast}^{t d_l} \, {V_{LCKM}}^{tb}} 
		= \left| \frac{{z_d}^{d_l b}}{
		{V_{LCKM}^\ast}^{t d_l} \, {V_{LCKM}}^{tb}}  \right| \, 
		e^{i \theta_{d_l}} \; , 
	\label{eqn:phasezdlvl}
\end{equation}
with
\begin{equation}
	\theta_{d_l} \equiv {\rm arg} \,  
		\left( \frac{{z_d}^{d_l b}}{
		{V_{LCKM}^\ast}^{t d_l} \, {V_{LCKM}}^{tb}} \right) \; .
	\label{eqn:thetadl}
\end{equation}

Each function in Eqs. (\ref{eqn:fl}) and (\ref{eqn:fr}) includes  the 
contributions of each diagram. That is, $F_{1,3}$, $F_2$, $F_{4,6}$ 
and $F_5$ come from the $W^\pm$, $\chi^\pm$, $Z$ and 
neutral Higgs ($H$, $\chi^0$) 
exchange diagrams respectively. These functions are given by
\begin{eqnarray}
	F_1 (x) & = & x \left[ 
		\frac{-113 + 151 \, x + 28 \, x^2}{72 \, (1 - x)^3} - 
		\frac{8 + 2 \, x - 21 \, x^2}{12 \, (1 - x)^4} \, \ln x
		\right] \; , 
		\label{eqn:f1} \\
	F_2 (x, m, m_i) & = & 
		4 \frac{1 - x^2 + 2 \, x \, \ln x}{3 \, (1 - x)^3} - 
		2 \, \frac{m_i}{m} 
		\frac{3 - 4 \, x + x^2 + 2 \, \ln x}{(1 - x)^3} \; , 
		\label{eqn:f2} \\
	F_3 (x) & = & x \left[ 
		\frac{25 - 53 \, x + 22 \, x^2}{36 \, (1 - x)^3} + 
		\frac{4 - 8 \, x + 3 \, x^2}{6 \, (1 - x)^4} \, \ln x
		\right] \; , 
		\label{eqn:f3} \\
	F_4 (x) & = & 
		\frac{-10 + 5 \, x - 11 \, x^2}{6 \, (1 - x)^3} - 
		\frac{3 - 4 \, x}{(1 - x)^4} \, x \, \ln x \; , 
		\label{eqn:f4} \\
	F_5 (x, y) & = & 
		x \, \left[
		\frac{-20 + 19 \, x - 5 \, x^2}{24 \, (1 - x)^3} - 
		\frac{2 - x}{4 \, (1 - x)^4} \, \ln x
		\right] 
		\nonumber \\
	& & 	- y \, \left[
		\frac{-16 + 29 \, y - 7 \, y^2}{24 \, (1 - y)^3} - 
		\frac{2 - 3 \, y}{4 \, (1 - y)^4} \, \ln y
		\right] \; ,
		\label{eqn:f5} \\
	F_6 (x) & = & 
		\frac{4 - 9 \, x - 5 \, x^3 - 6 \, x (1 - 2 \, x) \, \ln x}{
			12 \, (1 - x)^4} \; .
		\label{eqn:f6} 
\end{eqnarray}

From these results, one can predict the possible 
enhancements for the violations of $V-A$ structure as well as FCNC. 
The 
enhancement of the violations of $V-A$ structure is possibly large as there 
are some terms proportional to the third- and fourth-generation 
up-quark masses which are assumed to be large 
($m_{u^4} > m_t \sim 174$(GeV)). 
Also in the present model one can not neglect the 
left-handed term of the amplitude as is always done in the simplified 
SM calculations with same reason. 

\subsection{\bf CONSTRAINTS FOR THE MIXINGS AND MASSES}
\label{sec:constraint}

Before doing further evaluation, we impose experimental 
constraints on the mixing parameters and vector-like quark masses 
of Eqs. (\ref{eqn:fl}) and (\ref{eqn:fr}).
First, consider the mixing parameter ${z_d}^{d_l b}$ which indicates the 
FCNC in the $d_l b$ sector. Using the experimental results \cite{pdg, ua1}, 
one finds the upper-bounds for them as,
\begin{equation}
	\left| {z_d}^{d_l b} \right| \leq 1.7 \times {10}^{-3} \; , 
		\label{eqn:ubzddlb}
\end{equation}
from $Br (B \rightarrow X_{d_l} \, \mu^+ \, \mu^-) 
\leq 5 \times {10}^{-5}$ 
under an assumption that the Z-mediated tree-level Feynman
 diagram is dominant 
\cite{handoko}. For $i = t$ and $d_l = s$, one finds the ratio of mixing 
in Eqs. (\ref{eqn:fl}) and (\ref{eqn:fr}) as follows: 
\begin{equation}
	\left| \frac{{z_d}^{sb}}{
		{V_{LCKM}^\ast}^{ts} \, {V_{LCKM}}^{tb}}
	\right| \leq 4.0 \times {10}^{-2} \; ,
	\label{eqn:ratzv}
\end{equation}
from the assumption (A) of Sec.\ref{sec:model} 
for left-handed CKM matrix elements, 
$\left| {V_{LCKM}}^{ts} \right| \sim \left| {V_{LCKM}}^{cb} \right| 
\sim 0.042$ and $\left| {V_{LCKM}}^{tb} \right| \sim 1$ \cite{pdg}. 
We find that the contribution from the $z_d^{d^l b}(d^l \not= b)$ 
term is small enough to neglect. 

Our next task is determining the bound on the vector-like quark masses,
$m_{u^4}$ and $m_{d^4}$, and the mixings $\left| V^{td^i} \right|$ and 
$\left| V^{u^4 d^i} \right|$ which can be done by considering the experiment 
measurement of $B \rightarrow X_s \, \gamma$ process 
and the $T_{\rm new}$ parameter. 
The ratio of branching ratio is given as,
\begin{equation}
        \frac{Br(b \rightarrow d_l \, \gamma)}{
		Br(b \rightarrow c \, l \, \nu)} =  
                \frac{3 \, \alpha \, {Q_u}^2}{
			2 \, \pi \, f({m_c}^2/{m_b}^2)} \, 
		\left| \frac{{V_{LCKM}^\ast}^{t d_l} \, {V_{LCKM}}^{tb}}{
			{V_{LCKM}}^{cb}} \right|^2 \; \left( 
		\frac{{m_{d_l}}^2}{{m_b}^2} \, {F_L(m_{d_l})}^2 + 
					{F_R(m_{d_l})}^2 \right) \: .
		\label{eqn:br}
\end{equation}
which is normalized by the semileptonic decay $b \rightarrow c \, l \, \nu$ 
to reduce the uncertainties due to $m_b$ and the left CKM matrix element, 
especially for $d_l = s$. The phase space factor is
\begin{equation}
	f(y) \equiv 1 - 8 \, y + 8 \, y^3 - y^4 - 12 \, y^2 \, \ln y \: ,
	\label{eqn:fy} 
\end{equation}
and equals $0.520$ for ${m_c}/{m_b} = 0.3$ \cite{grig}. Note that the 
left handed term is kept because there is a possible enhancement due 
to the heavy quark mass from the ${m_i}/{m}$ term in Eq. (\ref{eqn:f2}). 
\begin{figure}[t]
        \begin{center}
\setlength{\unitlength}{0.240900pt}
\begin{picture}(1500,900)(0,0)
\tenrm
\thicklines \path(220,113)(240,113)
\thicklines \path(1436,113)(1416,113)
\put(198,113){\makebox(0,0)[r]{$0$}}
\thicklines \path(220,209)(240,209)
\thicklines \path(1436,209)(1416,209)
\put(198,209){\makebox(0,0)[r]{$0.005$}}
\thicklines \path(220,304)(240,304)
\thicklines \path(1436,304)(1416,304)
\put(198,304){\makebox(0,0)[r]{$0.010$}}
\thicklines \path(220,400)(240,400)
\thicklines \path(1436,400)(1416,400)
\put(198,400){\makebox(0,0)[r]{$0.015$}}
\thicklines \path(220,495)(240,495)
\thicklines \path(1436,495)(1416,495)
\put(198,495){\makebox(0,0)[r]{$0.020$}}
\thicklines \path(220,591)(240,591)
\thicklines \path(1436,591)(1416,591)
\put(198,591){\makebox(0,0)[r]{$0.025$}}
\thicklines \path(220,686)(240,686)
\thicklines \path(1436,686)(1416,686)
\put(198,686){\makebox(0,0)[r]{$0.030$}}
\thicklines \path(220,782)(240,782)
\thicklines \path(1436,782)(1416,782)
\put(198,782){\makebox(0,0)[r]{$0.035$}}
\thicklines \path(220,877)(240,877)
\thicklines \path(1436,877)(1416,877)
\put(198,877){\makebox(0,0)[r]{$0.040$}}
\thicklines \path(441,113)(441,133)
\thicklines \path(441,877)(441,857)
\put(441,68){\makebox(0,0){$0.001$}}
\thicklines \path(662,113)(662,133)
\thicklines \path(662,877)(662,857)
\put(662,68){\makebox(0,0){$0.002$}}
\thicklines \path(883,113)(883,133)
\thicklines \path(883,877)(883,857)
\put(883,68){\makebox(0,0){$0.003$}}
\thicklines \path(1104,113)(1104,133)
\thicklines \path(1104,877)(1104,857)
\put(1104,68){\makebox(0,0){$0.004$}}
\thicklines \path(1325,113)(1325,133)
\thicklines \path(1325,877)(1325,857)
\put(1325,68){\makebox(0,0){$0.005$}}
\thicklines \path(220,113)(1436,113)(1436,877)(220,877)(220,113)
\put(-25,495){\makebox(0,0)[l]{\shortstack{$\left| V^{ts} \right| $}}}
\put(828,0){\makebox(0,0){$\left| V^{tb} \right| $}}
\Thicklines \path(1354,113)(1354,113)(1352,151)(1350,189)(1345,228)(1339,266)(1332,304)(1321,342)(1310,380)(1295,419)(1279,457)(1259,495)(1235,533)(1210,571)(1182,610)(1146,648)(1109,686)(1065,724)(220,778)
\end{picture}
        \end{center}
	\caption{The allowed upper-bounds for $\left| V^{ts} \right|$ and 
		$\left| V^{tb} \right|$ from $b \rightarrow s \, \gamma$.}
	\label{fig:vrtsb}
\end{figure}
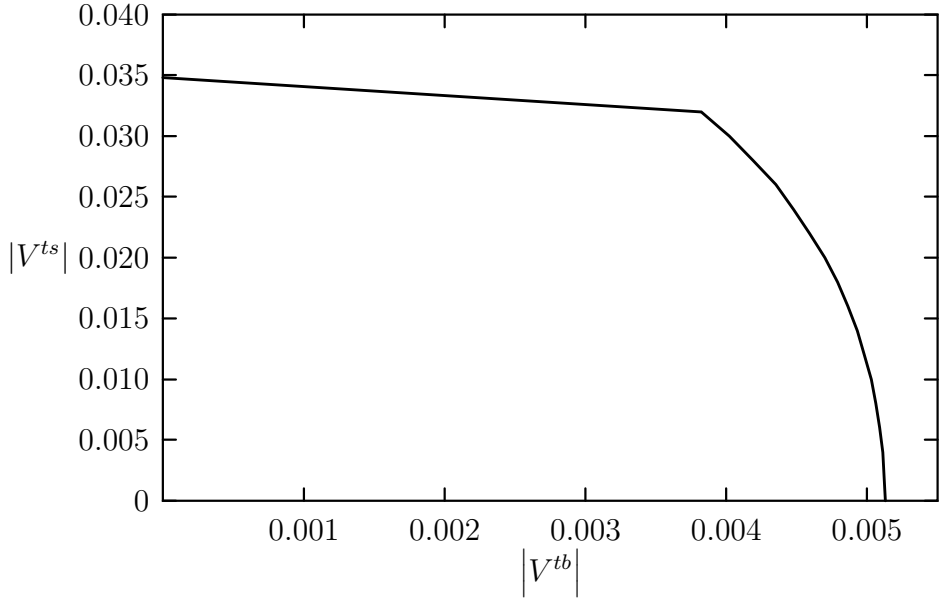

Now we can determine the upper-bounds for $\left| V^{ts} \right|$ 
and $\left| V^{tb} \right|$. Fortunately, the 
$b \rightarrow s \, \gamma$ process, the contribution of FCNC with the 
upper-bound in Eq. (\ref{eqn:ratzv}) is negligible \cite{handoko}.
By putting ${V_{LCKM}}^{u^4s} = 0$ at once and using the 
value, $Br (B \rightarrow X_s \, \gamma) = 5 \times {10}^{-4}$, 
$Br(B \rightarrow X_c \, l \, \nu) = 0.12$ and the maximum 
phase factors $\theta_s = \theta_{ts} = 0$, we find the 
upper-bounds as depicted in Fig. \ref{fig:vrtsb}. Here we find that the 
$\left| V^{tb} \right| \sim \left| V_R^{tb} \right|$ must be less than 0.0052.
These results are reasonable because 
the $b \rightarrow s \, \gamma$ process can be explained within the SM, 
hence the contribution of the new physics should be small. 

\begin{figure}[t]
	\begin{center}
\setlength{\unitlength}{0.240900pt}
\begin{picture}(1500,900)(0,0)
\tenrm
\thicklines \path(220,113)(240,113)
\thicklines \path(1436,113)(1416,113)
\put(198,113){\makebox(0,0)[r]{0}}
\thicklines \path(220,240)(240,240)
\thicklines \path(1436,240)(1416,240)
\put(198,240){\makebox(0,0)[r]{0.2}}
\thicklines \path(220,368)(240,368)
\thicklines \path(1436,368)(1416,368)
\put(198,368){\makebox(0,0)[r]{0.4}}
\thicklines \path(220,495)(240,495)
\thicklines \path(1436,495)(1416,495)
\put(198,495){\makebox(0,0)[r]{0.6}}
\thicklines \path(220,622)(240,622)
\thicklines \path(1436,622)(1416,622)
\put(198,622){\makebox(0,0)[r]{0.8}}
\thicklines \path(220,750)(240,750)
\thicklines \path(1436,750)(1416,750)
\put(198,750){\makebox(0,0)[r]{1.0q}}
\thicklines \path(220,877)(240,877)
\thicklines \path(1436,877)(1416,877)
\put(198,877){\makebox(0,0)[r]{1.2}}
\thicklines \path(220,113)(220,133)
\thicklines \path(220,877)(220,857)
\put(220,68){\makebox(0,0){0.50}}
\thicklines \path(342,113)(342,133)
\thicklines \path(342,877)(342,857)
\put(342,68){\makebox(0,0){0.52}}
\thicklines \path(463,113)(463,133)
\thicklines \path(463,877)(463,857)
\put(463,68){\makebox(0,0){0.54}}
\thicklines \path(585,113)(585,133)
\thicklines \path(585,877)(585,857)
\put(585,68){\makebox(0,0){0.56}}
\thicklines \path(706,113)(706,133)
\thicklines \path(706,877)(706,857)
\put(706,68){\makebox(0,0){0.58}}
\thicklines \path(828,113)(828,133)
\thicklines \path(828,877)(828,857)
\put(828,68){\makebox(0,0){0.60}}
\thicklines \path(950,113)(950,133)
\thicklines \path(950,877)(950,857)
\put(950,68){\makebox(0,0){0.62}}
\thicklines \path(1071,113)(1071,133)
\thicklines \path(1071,877)(1071,857)
\put(1071,68){\makebox(0,0){0.64}}
\thicklines \path(1193,113)(1193,133)
\thicklines \path(1193,877)(1193,857)
\put(1193,68){\makebox(0,0){0.66}}
\thicklines \path(1314,113)(1314,133)
\thicklines \path(1314,877)(1314,857)
\put(1314,68){\makebox(0,0){0.68}}
\thicklines \path(1436,113)(1436,133)
\thicklines \path(1436,877)(1436,857)
\put(1436,68){\makebox(0,0){0.70}}
\thicklines \path(220,113)(1436,113)(1436,877)(220,877)(220,113)
\put(-25,495){\makebox(0,0)[l]{\shortstack{$ T_{\rm new} $}}}
\put(828,0){\makebox(0,0){$\left| D_R^{34} \right|$}}
\Thicklines \path(220,228)(220,228)(271,238)(321,248)(372,258)
(423,269)(473,280)(524,292)(575,304)(625,317)(676,331)(727,345)
(777,359)(828,375)(879,390)(929,407)(980,424)(1031,442)(1081,460)
(1132,480)(1183,500)(1233,520)(1284,542)(1335,564)(1385,587)(1436,611)
\thinlines \path(220,277)(220,277)(271,291)(321,305)(372,320)
(423,335)(473,352)(524,369)(575,386)(625,405)(676,424)(727,444)
(777,465)(828,487)(879,510)(929,534)(980,559)(1031,584)(1081,611)
(1132,639)(1183,667)(1233,697)(1284,728)(1335,760)(1385,793)(1436,827)
\thinlines \dashline[-10]{25}(220,335)(220,335)(271,353)(321,372)(372,393)
(423,414)(473,436)(524,459)(575,483)(625,508)(676,535)(727,562)
(777,591)(828,620)(879,651)(929,684)(980,717)(1031,752)(1081,788)
(1132,826)(1183,865)(1198,877)
\Thicklines \dashline[-10]{25}(220,401)(220,401)(271,425)(321,450)(372,476)
(423,504)(473,533)(524,563)(575,594)(625,627)(676,662)(727,697)
(777,735)(828,774)(879,814)(929,856)(953,877)
\end{picture}
     	\end{center}
	\caption{The dependence of $T_{\rm new}$ on 
		$\left| D_R^{34} \right|$ for 
		$m_{d^4} = 250$ (GeV) (thick solid line), 
		$m_{d^4} = 300$ (GeV) (thin solid line), 
		$m_{d^4} = 350$ GeV) (thin dashed line) and 
		$m_{d^4} = 400$ GeV) (thick dashed line) with 
		$\left| {V_{LCKM}}^{td^4} \right| = 0.01 $.}
	\label{fig:tp}
\end{figure}
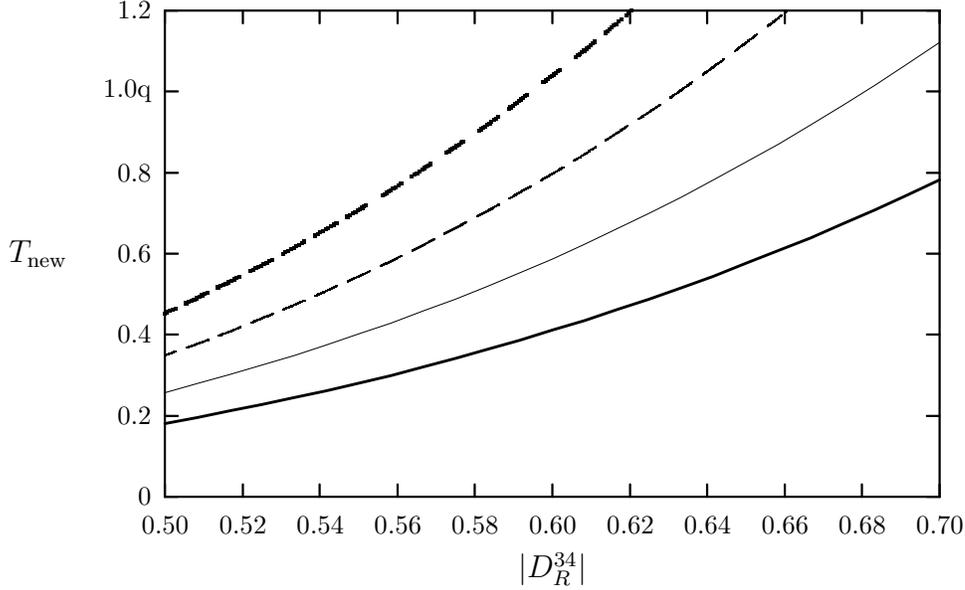

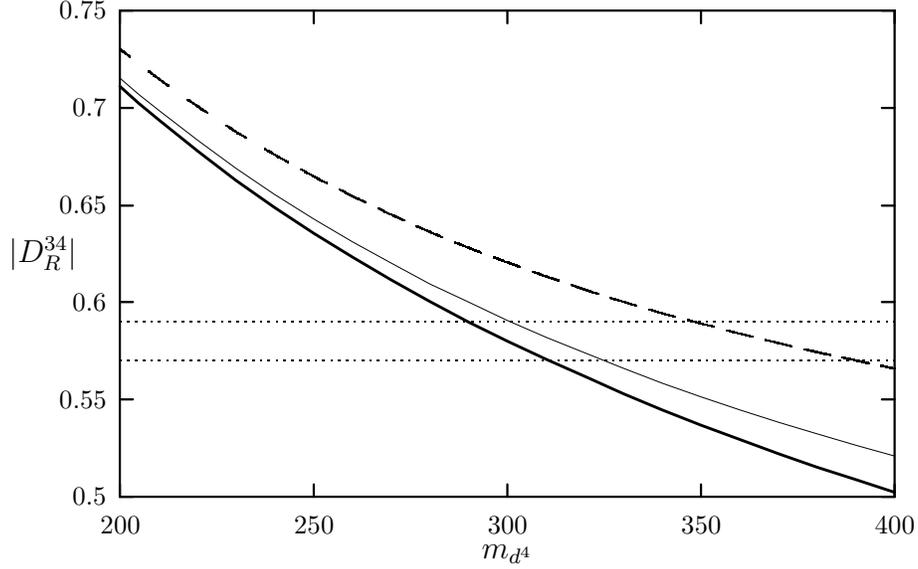
\begin{figure}[t]
	\begin{center}
\setlength{\unitlength}{0.240900pt}
\begin{picture}(1500,900)(0,0)
\tenrm
\thicklines \path(220,113)(240,113)
\thicklines \path(1436,113)(1416,113)
\put(198,113){\makebox(0,0)[r]{0.5}}
\thicklines \path(220,266)(240,266)
\thicklines \path(1436,266)(1416,266)
\put(198,266){\makebox(0,0)[r]{0.55}}
\thicklines \path(220,419)(240,419)
\thicklines \path(1436,419)(1416,419)
\put(198,419){\makebox(0,0)[r]{0.6}}
\thicklines \path(220,571)(240,571)
\thicklines \path(1436,571)(1416,571)
\put(198,571){\makebox(0,0)[r]{0.65}}
\thicklines \path(220,724)(240,724)
\thicklines \path(1436,724)(1416,724)
\put(198,724){\makebox(0,0)[r]{0.7}}
\thicklines \path(220,877)(240,877)
\thicklines \path(1436,877)(1416,877)
\put(198,877){\makebox(0,0)[r]{0.75}}
\thicklines \path(220,113)(220,133)
\thicklines \path(220,877)(220,857)
\put(220,68){\makebox(0,0){200}}
\thicklines \path(524,113)(524,133)
\thicklines \path(524,877)(524,857)
\put(524,68){\makebox(0,0){250}}
\thicklines \path(828,113)(828,133)
\thicklines \path(828,877)(828,857)
\put(828,68){\makebox(0,0){300}}
\thicklines \path(1132,113)(1132,133)
\thicklines \path(1132,877)(1132,857)
\put(1132,68){\makebox(0,0){350}}
\thicklines \path(1436,113)(1436,133)
\thicklines \path(1436,877)(1436,857)
\put(1436,68){\makebox(0,0){400}}
\thicklines \path(220,113)(1436,113)(1436,877)(220,877)(220,113)
\put(45,495){\makebox(0,0)[l]{\shortstack{$ \left| D_R^{34} \right| $}}}
\put(828,23){\makebox(0,0){$ m_{d^4} $}}
\Thicklines \path(220,758)(220,758)(250,731)(281,705)(342,656)(402,610)
(463,568)(524,527)(585,490)(646,454)(706,420)(767,388)(828,358)(889,329)
(950,302)(1010,275)(1071,250)(1132,226)(1193,203)(1254,181)(1314,160)
(1375,140)(1436,120)
\thinlines \path(220,771)(220,771)(250,745)(281,720)(342,673)(402,629)
(463,588)(524,550)(585,514)(646,480)(706,448)(767,419)(828,390)(889,364)
(950,339)(1010,315)(1071,292)(1132,270)(1193,250)(1254,230)(1314,212)
(1375,194)(1436,177)
\thinlines \dashline[-10]{25}(220,816)(220,816)(281,769)
(342,726)(402,686)(463,650)
(524,616)(585,585)(646,556)(706,529)(767,504)(828,481)(889,459)(950,438)
(1010,419)(1071,401)(1132,385)(1193,369)(1254,354)(1314,340)(1375,327)
(1436,314)
\thinlines \dashline[-10]{3}(220,327)(220,327)(232,327)
(245,327)(257,327)(269,327)
(281,327)(294,327)(306,327)(318,327)(331,327)(343,327)(355,327)(367,327)
(380,327)(392,327)(404,327)(417,327)(429,327)(441,327)(453,327)(466,327)
(478,327)(490,327)(503,327)(515,327)(527,327)(539,327)(552,327)(564,327)
(576,327)(588,327)(601,327)(613,327)(625,327)(638,327)(650,327)(662,327)
(674,327)(687,327)(699,327)(711,327)(724,327)(736,327)(748,327)(760,327)
(773,327)(785,327)(797,327)(810,327)(822,327)
\thinlines \dashline[-10]{3}(822,327)(834,327)(846,327)(859,327)
(871,327)(883,327)
(896,327)(908,327)(920,327)(932,327)(945,327)(957,327)(969,327)(982,327)
(994,327)(1006,327)(1018,327)(1031,327)(1043,327)(1055,327)(1068,327)
(1080,327)(1092,327)(1104,327)(1117,327)(1129,327)(1141,327)(1153,327)
(1166,327)(1178,327)(1190,327)(1203,327)(1215,327)(1227,327)(1239,327)
(1252,327)(1264,327)(1276,327)(1289,327)(1301,327)(1313,327)(1325,327)
(1338,327)(1350,327)(1362,327)(1375,327)(1387,327)(1399,327)(1411,327)
(1424,327)(1436,327)
\thinlines \dashline[-10]{3}(220,388)(220,388)(232,388)
(245,388)(257,388)
(269,388)(281,388)(294,388)(306,388)(318,388)(331,388)(343,388)
(355,388)(367,388)(380,388)(392,388)(404,388)(417,388)(429,388)
(441,388)(453,388)(466,388)(478,388)(490,388)(503,388)(515,388)
(527,388)(539,388)(552,388)(564,388)(576,388)(588,388)(601,388)
(613,388)(625,388)(638,388)(650,388)(662,388)(674,388)(687,388)
(699,388)(711,388)(724,388)(736,388)(748,388)(760,388)(773,388)
(785,388)(797,388)(810,388)(822,388)
\thinlines \dashline[-10]{3}(822,388)(834,388)(846,388)(859,388)(871,388)
(883,388)(896,388)(908,388)(920,388)(932,388)(945,388)(957,388)
(969,388)(982,388)(994,388)(1006,388)(1018,388)(1031,388)(1043,388)
(1055,388)(1068,388)(1080,388)(1092,388)(1104,388)(1117,388)
(1129,388)(1141,388)(1153,388)(1166,388)(1178,388)(1190,388)
(1203,388)(1215,388)(1227,388)(1239,388)(1252,388)(1264,388)
(1276,388)(1289,388)(1301,388)(1313,388)(1325,388)(1338,388)
(1350,388)(1362,388)(1375,388)(1387,388)(1399,388)(1411,388)
(1424,388)(1436,388)
\end{picture}
     	\end{center}
	\caption{The upper bound of $\left| D_R^{34} \right|$ 
                and $m_{d^4}$  from the bound of $T_{NEW}$ for 
		$\left| {V_{LCKM}}^{td^4} \right| = 0$ (thick line), 
		$\left| {V_{LCKM}}^{td^4} \right| = 0.01$ (thin line), 
		$\left| {V_{LCKM}}^{td^4} \right| = 0.02$ (dashed line).
                The dotted lines show the bound,eq.(27) from $R_b$.}
	\label{fig:mvlq2}
\end{figure}

\begin{figure}[t]
	\begin{center}
\setlength{\unitlength}{0.240900pt}
\begin{picture}(1500,900)(0,0)
\tenrm
\thicklines \path(220,113)(240,113)
\thicklines \path(1436,113)(1416,113)
\put(198,113){\makebox(0,0)[r]{0}}
\thicklines \path(220,266)(240,266)
\thicklines \path(1436,266)(1416,266)
\put(198,266){\makebox(0,0)[r]{0.005}}
\thicklines \path(220,419)(240,419)
\thicklines \path(1436,419)(1416,419)
\put(198,419){\makebox(0,0)[r]{0.01}}
\thicklines \path(220,571)(240,571)
\thicklines \path(1436,571)(1416,571)
\put(198,571){\makebox(0,0)[r]{0.015}}
\thicklines \path(220,724)(240,724)
\thicklines \path(1436,724)(1416,724)
\put(198,724){\makebox(0,0)[r]{0.02}}
\thicklines \path(220,877)(240,877)
\thicklines \path(1436,877)(1416,877)
\put(198,877){\makebox(0,0)[r]{0.025}}
\thicklines \path(220,113)(220,133)
\thicklines \path(220,877)(220,857)
\put(220,68){\makebox(0,0){200}}
\thicklines \path(524,113)(524,133)
\thicklines \path(524,877)(524,857)
\put(524,68){\makebox(0,0){250}}
\thicklines \path(828,113)(828,133)
\thicklines \path(828,877)(828,857)
\put(828,68){\makebox(0,0){300}}
\thicklines \path(1132,113)(1132,133)
\thicklines \path(1132,877)(1132,857)
\put(1132,68){\makebox(0,0){350}}
\thicklines \path(1436,113)(1436,133)
\thicklines \path(1436,877)(1436,857)
\put(1436,68){\makebox(0,0){400}}
\thicklines \path(220,113)(1436,113)(1436,877)(220,877)(220,113)
\put(45,495){\makebox(0,0)[l]{\shortstack{$ \left| U_R^{34} \right|$}}}
\put(828,23){\makebox(0,0){$ m_{d^4} $}}
\Thicklines  \path(220,593)(220,593)(271,593)(321,593)(372,593)
(423,593)(473,593)(524,593)(575,593)(625,593)(676,593)(727,593)
(777,593)(828,593)(879,593)(929,593)(980,593)(1031,593)(1081,593)
(1132,593)(1183,593)(1233,593)(1284,593)(1335,593)(1385,593)(1436,593)
\Thicklines \dashline[-10]{25}(220,447)(220,447)(271,441)(321,435)(372,429)
(423,423)(473,417)(524,411)(575,405)(625,399)(676,393)(727,387)
(777,381)(828,374)(879,368)(929,362)(980,356)(1031,350)(1081,344)
(1132,338)(1183,332)(1233,326)(1284,320)(1335,314)(1385,308)(1436,302)
\thinlines \path(220,302)(220,302)(271,289)(321,277)(372,265)
(423,253)(473,241)(524,229)(575,216)(625,204)(676,192)(727,180)
(777,168)(828,156)(879,144)(929,131)(980,119)(1004,113)(1059,113)
(1071,118)(1132,141)(1193,164)(1254,187)(1314,210)(1375,232)(1436,254)
\thinlines \dashline[-10]{25}(220,113)(220,113)(281,113)(342,113)(402,113)
(425,113)(463,141)(524,185)(585,228)(646,270)(706,312)(767,353)
(828,393)(889,433)(950,472)(1010,511)(1071,549)(1132,587)(1193,624)
(1254,661)(1314,698)(1375,735)(1436,771)
\thinlines \dashline[-10]{3}(220,394)(220,394)(232,394)(245,394)(257,394)
(269,394)(281,394)(294,394)(306,394)(318,394)(331,394)(343,394)
(355,394)(367,394)(380,394)(392,394)(404,394)(417,394)(429,394)
(441,394)(453,394)(466,394)(478,394)(490,394)(503,394)(515,394)
(527,394)(539,394)(552,394)(564,394)(576,394)(588,394)(601,394)
(613,394)(625,394)(638,394)(650,394)(662,394)(674,394)(687,394)
(699,394)(711,394)(724,394)(736,394)(748,394)(760,394)(773,394)
(785,394)(797,394)(810,394)(822,394)
\thinlines \dashline[-10]{3}(822,394)(834,394)(846,394)(859,394)(871,394)
(883,394)(896,394)(908,394)(920,394)(932,394)(945,394)(957,394)
(969,394)(982,394)(994,394)(1006,394)(1018,394)(1031,394)(1043,394)
(1055,394)(1068,394)(1080,394)(1092,394)(1104,394)(1117,394)
(1129,394)(1141,394)(1153,394)(1166,394)(1178,394)(1190,394)
(1203,394)(1215,394)(1227,394)(1239,394)(1252,394)(1264,394)
(1276,394)(1289,394)(1301,394)(1313,394)(1325,394)(1338,394)
(1350,394)(1362,394)(1375,394)(1387,394)(1399,394)(1411,394)
(1424,394)(1436,394)
\end{picture}
     	\end{center}
	\caption{The lower bound of $\left| U_R^{34} \right|$ 
                from the bound $\left| D_R^{34} \right| > 0.55$ and 
                the upper bound of $T_{NEW}$ for 
		$\left| {V_{LCKM}}^{td^4} \right| = 0$ (thick line),
                $\left| {V_{LCKM}}^{td^4} \right| = 0.005$ (thick dashed line), 
		$\left| {V_{LCKM}}^{td^4} \right| = 0.01$ (thin line), 
		$\left| {V_{LCKM}}^{td^4} \right| = 0.02$ (thin dashed line). 
                The dotted line show the upper bound from $R_b$ and $
                b\rightarrow s \gamma $(Eq. (45)).}
	\label{fig:mvlq1}
\end{figure}
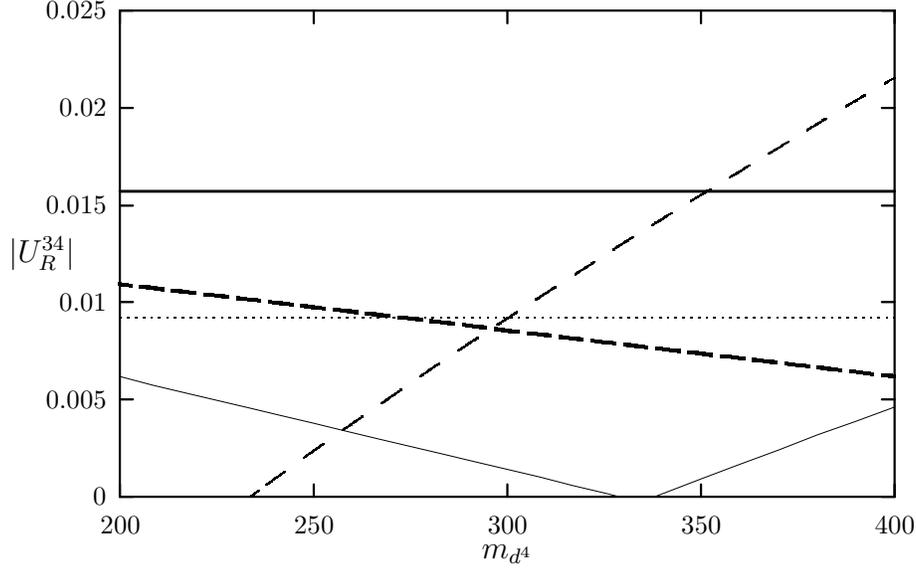

We find that the upper bound of $\left| V_R^{tb} \right|$ from $b \rightarrow 
s\gamma $.  
Hence, we can find  the upper bound of $\left| U_R^{34} \right|$ 
by using the relation for the right CKM matrix (\ref{eqn:vrckm})
and these bounds from $b \rightarrow s\gamma $ and $R_b$(\ref{Dbound}). 
\beq
	\left| U_R^{34} \right| < 0.0092 .
\eeq

Lastly, we are going to use $T_{\rm new}$ to determine the allowed 
regions of $\left| {U_R}^{34} \right|$. 
The interactions between the 
gauge bosons and the quarks which contribute to the $T$ parameter are,  
\begin{eqnarray}
	{\cal L}_{W^{1,3}} & = & \frac{g}{2} \left\{
	\pm \bar{q}^\alpha \, \gamma^\mu \, W_\mu^3 \, \left[ 
	\delta^{\alpha \beta} \, L + \left( \delta^{\alpha \beta} - 
		{z_q}^{\alpha \beta} \right) \, R \right] \, q^\beta 
	\right. \nonumber \\
	& & \left. 
	- \bar{u}^\alpha \, \gamma^\mu \, W_\mu^1 \, 
		{V_{LCKM}}^{\alpha \beta} \, \left( L + 
		V^{\alpha \beta} \, R \right) \, d^\beta + 
		{\rm h.c.} \right\} \; . 
	\label{eqn:lw13}	
\end{eqnarray}
The $T$ parameter becomes
\begin{eqnarray}
	T & = & \frac{N_c}{
		32 \, \pi \, {M_Z}^2 \, \sin^2 \theta_W \, \cos^2 \theta_W}
		\sum_{\alpha, \beta}^4 \left\{
	2 \, \left| {V_{LCKM}}^{\alpha \beta} \right|^2 \, 
		f(m_{u^\alpha}, m_{d^\beta}) \left( 
		1 + \left| V^{\alpha \beta} \right|^2 \right) 
		\right.
		\nonumber \\
	& & \left. 
	- 16 \, g(m_{u^\alpha}, m_{d^\beta}) \, {V_{LCKM}}^{\alpha \beta} \, 
		{\rm Re} V^{\alpha \beta} 
		\right.
		\nonumber \\
	& & \left. 
	- f(m_{u^\alpha}, m_{u^\beta}) \left[ \delta^{\alpha \beta} + 
		\left( \delta^{\alpha \beta} - 
		{z_u}^{\alpha \beta} \right)^2 \right] 
	- f(m_{d^\alpha}, m_{d^\beta}) \left[ \delta^{\alpha \beta} + 
		\left( \delta^{\alpha \beta} - 
		{z_d}^{\alpha \beta} \right)^2 \right] 
		\right.
		\nonumber \\
	& & \left. 
	- 8 \, h(m_{u^\alpha}) \, \left| {U_R}^{\alpha 4} \right|^2
	- 8 \, h(m_{d^\alpha}) \, \left| {D_R}^{\alpha 4} \right|^2
	\right\} \; ,  
	\label{eqn:tp}
\end{eqnarray}
from the definitions of $T$ parameter presented by Ref.\cite{peskin}. 
$N_c$ is the number of color. These auxiliary functions are defined as, 
\begin{eqnarray}
	f(x, y) & \equiv & x^2 + y^2 - 
		2 \, \frac{x^4 \, \ln x^2 - y^4 \, \ln y^2}{x^2 - y^2}
		\; , \\
	g(x, y) & \equiv & x \, y \, \left( 1 - \frac{
		x^2 \, \ln x^2 - y^2 \, \ln y^2}{x^2 - y^2}
		\right) \; , \\
	h(x)    & \equiv & x^2 \, \ln x^2 
		\; .
\end{eqnarray}
The $T_{\rm new}$ parameter is found by subtracting the SM contribution. 
According to Eqs. (\ref{eqn:f}), (\ref{eqn:vlckm}) and (\ref{eqn:vrckm}), 
one finds the following mass relations, 
\begin{eqnarray}
	\delta^{\alpha \beta} \, {m_d}^\beta \, {D_R}^{\beta 4} \, 
		{D_R^\dagger}^{4 \sigma} & = & 
	{V_{LCKM}^\dagger}^{\alpha \beta} \, {m_u}^\beta \, 
		{V_{RCKM}}^{\beta \sigma} \; , 
	\label{eqn:md} \\
	\delta^{\alpha \beta} \, {m_u}^\beta \, {U_R}^{\beta 4} \, 
		{U_R^\dagger}^{4 \sigma} & = & 
	{V_{LCKM}}^{\alpha \beta} \, {m_d}^\beta \, 
		{V_{RCKM}^\dagger}^{\beta \sigma} \; . 
	\label{eqn:mu}
\end{eqnarray}
Both equations yield a relation between $m_{u^4}$ and $m_{d^4}$, and
$U_R^{34}( \sim \sin \theta_U )$ and $ D_R^{34} (\sim \sin \theta_D)$ under the
assumptions (A) and (B) ( Eq.(22) and Eq.(23) ).
\bea
	m_{u^4} &=& \frac{1}{\cos \theta_U} \, \left( 
		-m_b \, {V_{LCKM}}^{td^4} \, \sin \theta_D + 
		m_{d^4} \, \cos \theta_D \right) \; ,
	\label{eqn:mu4md4}\\
        \sin \theta_U &=& \frac{m_b}{m_t} \sin \theta_D 
                     - \frac{m_{d^4}}{m_t} {V_{LCKM}}^{td^4} \cos \theta_D 
        \label{eqn:susd}
\eea
Using these relations, we can simplify our expression for $T_{new}$ 
and plot the figures.
Fig.3 shows $T_{new}$ {\em vs} $\left| D_R^{34} \right| $. Fig.4 shows 
the upper bound on $\left| D_R^{34} \right|$ and $m_{d^4}$ from the upper 
bound of 
$T_{new}$, ($T_{new} < 0.55 $ for $m_t = 175GeV$ and 
$m_H = 1 TeV$)\cite{pdg,hagi}. By using the eq. (54), we obtain 
the lower bound on 
$\left| U_R^{34} \right|$ from the lower bound of $\left| D_R^{34} \right|
( > 0.55 )$ that is the bound when there is the deviation from the 
SM prediction in $R_b$.(See the Fig.1.)   
In the figures we put  
$m_t = 175$ (GeV), $m_b = 5$ (GeV) and 
$\left| {V_{LCKM}}^{td^4} \right| \sim \left| {V_{LCKM}}^{u^4 b} \right|$, 
and also neglect tiny contributions from the mixings between 
light-quarks and the fourth generation quarks. 

Unfortunately, after calculating $T_{\rm new}$ we find that there 
is no allowed region 
for $T_{\rm new} < 0.55$ for any values 
of vector-like quark masses 
($m_{u^4}$ and $m_{d^4}$) consistent with
 assumption (A) of Sec.\ref{sec:model} of left-handed CKM matrix. 
When $\left| {V_{LCKM}^{td^4}} \right| = 0$, Fig. \ref{fig:mvlq2} shows that 
$m_d^4$ is less than about 300GeV, and 
Fig. \ref{fig:mvlq1} shows that there is no region to satisfy the bound of 
$\left| U_R^{34} \right|$, Eq. (45).  
Therefore, let us use more general of the LCKM matrix for further analysis. 
If $\left| {V_{LCKM}}^{td^4} \right| \not=0$, in Figs. 
\ref{fig:mvlq2} and \ref{fig:mvlq1}, the allowed regions 
for vector-like quark mass $m_{d^4}$ for some values 
of $\left| U_R^{34} \right|$, 
$\left| D_R^{34} \right|$ and $\left| {V_{LCKM}}^{td^4} \right|$ 
are plotted by using 
Eqs. (\ref{eqn:tp}), (\ref{eqn:mu4md4}) and (\ref{eqn:susd}).

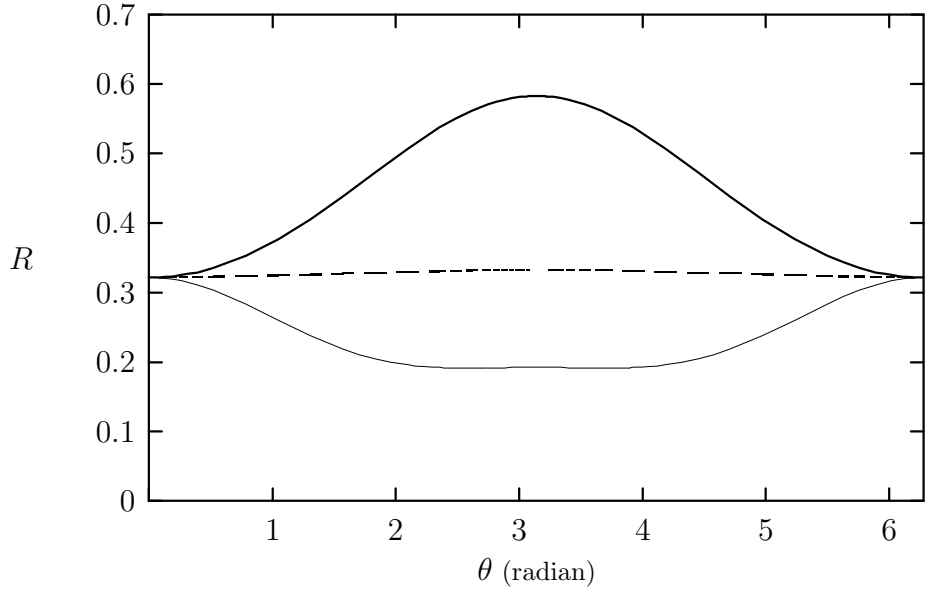
\begin{figure}[t]
	\begin{center}
\setlength{\unitlength}{0.240900pt}
\begin{picture}(1500,900)(0,0)
\tenrm
\thicklines \path(220,113)(240,113)
\thicklines \path(1436,113)(1416,113)
\put(198,113){\makebox(0,0)[r]{$0$}}
\thicklines \path(220,222)(240,222)
\thicklines \path(1436,222)(1416,222)
\put(198,222){\makebox(0,0)[r]{$0.1$}}
\thicklines \path(220,331)(240,331)
\thicklines \path(1436,331)(1416,331)
\put(198,331){\makebox(0,0)[r]{$0.2$}}
\thicklines \path(220,440)(240,440)
\thicklines \path(1436,440)(1416,440)
\put(198,440){\makebox(0,0)[r]{$0.3$}}
\thicklines \path(220,550)(240,550)
\thicklines \path(1436,550)(1416,550)
\put(198,550){\makebox(0,0)[r]{$0.4$}}
\thicklines \path(220,659)(240,659)
\thicklines \path(1436,659)(1416,659)
\put(198,659){\makebox(0,0)[r]{$0.5$}}
\thicklines \path(220,768)(240,768)
\thicklines \path(1436,768)(1416,768)
\put(198,768){\makebox(0,0)[r]{$0.6$}}
\thicklines \path(220,877)(240,877)
\thicklines \path(1436,877)(1416,877)
\put(198,877){\makebox(0,0)[r]{$0.7$}}
\thicklines \path(220,113)(220,133)
\thicklines \path(414,113)(414,133)
\thicklines \path(414,877)(414,857)
\put(414,68){\makebox(0,0){$1$}}
\thicklines \path(607,113)(607,133)
\thicklines \path(607,877)(607,857)
\put(607,68){\makebox(0,0){$2$}}
\thicklines \path(801,113)(801,133)
\thicklines \path(801,877)(801,857)
\put(801,68){\makebox(0,0){$3$}}
\thicklines \path(995,113)(995,133)
\thicklines \path(995,877)(995,857)
\put(995,68){\makebox(0,0){$4$}}
\thicklines \path(1188,113)(1188,133)
\thicklines \path(1188,877)(1188,857)
\put(1188,68){\makebox(0,0){$5$}}
\thicklines \path(1382,113)(1382,133)
\thicklines \path(1382,877)(1382,857)
\put(1382,68){\makebox(0,0){$6$}}
\thicklines \path(220,113)(1436,113)(1436,877)(220,877)(220,113)
\put(0,495){\makebox(0,0)[l]{\shortstack{$R$}}}
\put(828,0){\makebox(0,0){$\theta$ (radian)}}
\thinlines \dashline[-10]{25}(220,464)(220,464)(222,464)(223,464)
(225,464)(226,464)(228,464)(230,464)(233,464)(236,464)(239,464)
(245,464)(252,464)(258,464)(271,464)(296,464)(321,465)(372,466)
(423,467)(473,468)(524,470)(575,471)(626,472)(676,474)(702,474)
(727,475)(752,475)(765,475)(778,475)(790,475)(797,475)(803,475)
(809,475)(812,475)(816,475)(819,475)(820,475)(822,475)(824,475)
(825,475)(827,475)(828,475)(830,475)(831,475)(833,475)(835,475)
(836,475)(838,475)(841,475)(844,475)(847,475)
\thinlines \dashline[-10]{25}(847,475)(854,475)(860,475)(866,475)
(879,475)(904,475)(930,475)(980,474)(1031,472)(1082,471)(1132,470)
(1183,468)(1234,467)(1285,466)(1310,465)(1335,465)(1361,464)
(1386,464)(1399,464)(1405,464)(1411,464)(1418,464)(1421,464)
(1424,464)(1427,464)(1429,464)(1430,464)(1432,464)(1433,464)
(1435,464)(1436,464)
\thinlines \path(220,464)(220,464)(222,464)(223,464)(225,464)
(226,464)(228,464)(230,464)(233,464)(236,463)(239,463)(245,463)
(258,461)(271,459)(296,452)(321,444)(372,422)(423,397)(473,373)
(524,352)(550,343)(575,336)(600,331)(626,327)(638,325)(651,324)
(664,323)(670,323)(676,323)(683,322)(689,322)(695,322)(698,322)
(702,322)(705,322)(708,322)(709,322)(711,322)(713,322)(714,322)
(716,322)(717,322)(719,322)(721,322)(722,322)(724,322)(725,322)
(727,322)(733,322)(736,322)(740,322)
\thinlines \path(740,322)(746,322)(752,322)(778,322)(790,323)
(803,323)(809,323)(812,323)(816,323)(819,323)(820,323)(822,323)
(824,323)(825,323)(827,323)(828,323)(830,323)(831,323)(833,323)
(835,323)(836,323)(838,323)(841,323)(847,323)(854,323)(866,323)
(879,322)(904,322)(911,322)(917,322)(920,322)(923,322)(927,322)
(928,322)(930,322)(931,322)(933,322)(934,322)(936,322)(938,322)
(939,322)(941,322)(942,322)(944,322)(946,322)(949,322)(952,322)
(955,322)(961,322)(968,322)(974,322)
\thinlines \path(974,322)(980,323)(993,323)(1006,324)(1018,325)
(1031,327)(1056,331)(1082,336)(1107,343)(1132,352)(1183,373)
(1234,397)(1285,422)(1310,434)(1335,444)(1361,452)(1386,459)
(1399,461)(1405,462)(1411,463)(1418,463)(1421,463)(1424,464)
(1427,464)(1429,464)(1430,464)(1432,464)(1433,464)(1435,464)(1436,464)
\thicklines \path(220,464)(220,464)(222,464)(223,464)(225,464)
(226,464)(228,464)(230,464)(233,464)(236,464)(239,464)(245,465)
(252,465)(258,466)(271,468)(283,470)(296,472)(321,479)(372,498)
(423,524)(473,555)(524,591)(575,629)(626,666)(676,700)(702,714)
(727,726)(752,736)(765,740)(778,743)(790,746)(797,747)(803,748)
(809,748)(812,748)(816,749)(819,749)(820,749)(822,749)(824,749)
(825,749)(827,749)(828,749)(830,749)(831,749)(833,749)(835,749)
(836,749)(838,749)(841,749)(844,748)
\thicklines \path(844,748)(847,748)(854,748)(860,747)(866,746)
(879,743)(904,736)(930,726)(980,700)(1031,666)(1082,629)(1132,591)
(1183,555)(1234,524)(1285,498)(1310,488)(1335,479)(1361,472)
(1373,470)(1386,468)(1399,466)(1405,465)(1411,465)(1418,464)
(1421,464)(1424,464)(1427,464)(1429,464)(1430,464)(1432,464)
(1433,464)(1435,464)(1436,464)
\end{picture}
        \end{center}
        \caption{$R$ as a function of $\theta_{u^4 b}$ (thick line) 
		with $\theta_d = \theta_{u^4 d} = 0$, 
		$\theta_{d}$ (dashed line) with $\theta_{u^4 b} = 
		\theta_{u^4 d} = 0$ and $\theta_{u^4 d}$ 
		(thin line) with $\theta_{u^4 b} = \theta_d = 0$.}
	\label{fig:r}
\end{figure}

\subsection{\bf EFFECTS IN THE RATIO $R$}
\label{sec:result}

Now we are ready to evaluate the contributions of the FCNC's and 
the violation of $V-A$ structure to the ratio $R$. Note 
that, in the present paper, the QCD corrections are not included. 
$R$ is defined as
\begin{equation}
	R \equiv \frac{Br(b \rightarrow d \, \gamma)
		}{Br(b \rightarrow s \, \gamma)} \: . 
		\label{eqn:r}
\end{equation}
Then, from Eq. (\ref{eqn:br}),
\begin{equation}
	R = \left| \frac{{V_{LCKM}}^{td}}{{V_{LCKM}}^{ts}} \right|^2 \, 
		\frac{\left( {{m_d}^2}/{{m_b}^2}\right) \, {F_L(m_d)}^2 
			+ {F_R(m_d)}^2}{
		\left( {{m_s}^2}/{m_b}^2 \right) \, {F_L(m_s)}^2 + 
			{F_R(m_s)}^2} \: .
		\label{eqn:ratio}
\end{equation}
The type $B \rightarrow X_d \, \gamma$ decay has not been observed 
yet, but upper-bounds have been obtained from the exclusive decays 
$B \rightarrow (\rho^-, \rho^0, \omega) \, \gamma$ \cite{poling}.
Then, within the SM,  
\begin{equation}
	\left| \frac{{V_{LCKM}}^{td}}{{V_{LCKM}}^{ts}} \right| < 
		\frac{1}{1.8} \; ,
	\label{eqn:vtdvts}
\end{equation} 
has been extracted, and theoretically the ratio becomes $R < 0.31$.
We have to include all of FCNC's 
effects because the contributions are not negligible \cite{handoko}.
However, under the assumption and experimental bounds in the preceding 
section, the terms $\left| {D_R}^{i4} \right|^2$ can be neglected safely, 
since they are small compared to the diagonal ones.  

According to the results of the preceding section, for further numerical 
calculations, we put $\left| D_R^{34} \right| = 0.58$, 
$\left| U_R^{34} \right| = 0.004$, $m_{d^4} = 290$ (GeV), 
$\sin^2 \theta_W = 0.234$, $M_H = 500$ (GeV) and  
$m_t = 174$ (GeV), $m_b = 5$ (GeV).
With these values, one finds respectively 
$\left| {D_R}^{44} \right| \sim 0.815$,  
$\left| V^{tb} \right| = 0.002$, $\left| V^{u^4 b} \right| = 0.006$, 
$\left| V^{ts} \right| \sim 0.033$ and $\left| V^{u^4 s} \right| \sim 3.7$.  
On the other hand, rough order estimations by considering unitarity of 
$V_{LCKM}$ and Eq. (\ref{eqn:ass}), gives 
$\left| V^{u^4 d} \right| \sim 0.01$, 
$\left| {{V_{LCKM}^\ast}^{u^4 d} \, {V_{LCKM}}^{u^4 b}}/{
{V_{LCKM}^\ast}^{td}} \right| \sim 0.01$, $\left| V^{td} \right| \sim 0.01$ 
and  $\left| {{z_d}^{db}}/{{V_{LCKM}}^{td}} \right| \sim 0.1$.
In Fig. \ref{fig:r}, the ratio is depicted as a function of the 
phases, $\theta_d$, $\theta_{u^4 d}$ and $\theta_{u^4 b}$. 
On the other hand, since the contributions from the terms of $V^{td}$ 
and $V^{tb}$ are comparably small, the deviations due to its phase 
should be negligible.

\subsection{\bf CONCLUSIONS}
\label{sec:conclusion}

The effects of extending the SM by including a vector-like doublet 
of quarks on the FCNC and the violation of $V-A$ structure on the ratio 
$R$ has been studied. Including the constraints 
from experimental results, we found that there will be significant 
deviations in $R$ even for small mixings. 

To satisfy the constraints from $R_b$, $b\rightarrow s\gamma$ and the $T$ 
parameter on the vector-like doublet model, $m_{d^4}$ must be less than 
about 350 (GeV) and
$\left| V_{LCKM}^{td^4} \right| \not= 0$. However an allowed region 
still exists even if $\left| D_{RCKM}^{34} \right|$ has a large value.  

Lastly, because the 
experimental constraint of Eq. (\ref{eqn:vtdvts}) is not strong, 
improved data and the eventual detection of the type 
$B \rightarrow X_d \, \gamma$ decays will provide an important test of 
the SM and will further constrain the 
present model. 

\bigskip
\noindent
{\Large \bf ACKNOWLEDGMENT}

We would like to thank Dr.R.Szalapski for reading the manuscript and 
some kindly comments.
L.T.H. expresses his gratitude to the Japanese Government for financially 
supporting his work under Monbusho Fellowship.

\end{document}